# XeMap: Contextual Referring in Large-Scale Remote Sensing Environments

Yuxi Li, Lu Si, Yujie Hou, Chengaung Liu, Bin Li, Hongjian Fang, and Jun Zhang

*Abstract*—Advancements in remote sensing (RS) imagery have provided high-resolution detail and vast coverage, yet existing methods, such as image-level captioning/retrieval and object-level detection/segmentation, often fail to capture mid-scale semantic entities essential for interpreting large-scale scenes. To address this, we propose the conte*X*tual r*e*ferring *Map* (*XeMap*) task, which focuses on contextual, fine-grained localization of text-referred regions in large-scale RS scenes. Unlike traditional approaches, XeMap enables precise mapping of mid-scale semantic entities that are often overlooked in image-level or object-level methods. To achieve this, we introduce XeMap-Network, a novel architecture designed to handle the complexities of pixel-level cross-modal contextual referring mapping in RS. The network includes a fusion layer that applies self- and cross-attention mechanisms to enhance the interaction between text and image embeddings. Furthermore, we propose a Hierarchical Multi-Scale Semantic Alignment (HMSA) module that aligns multiscale visual features with the text semantic vector, enabling precise multimodal matching across large-scale RS imagery. To support XeMap task, we provide a novel, annotated dataset, XeMap-set, specifically tailored for this task, overcoming the lack of XeMap datasets in RS imagery. XeMap-Network is evaluated in a zero-shot setting against state-of-the-art methods, demonstrating superior performance. This highlights its effectiveness in accurately mapping referring regions and providing valuable insights for interpreting large-scale RS environments.

## I. INTRODUCTION

LARGE-SCALE remote sensing (RS) imagery, empowered by recent advances in imaging technology such as array cameras [1]-[3] and high-resolution electro-optical sensors [4], can now provide high-resolution detail information while covering vast areas up to kilometer-scale [5]. Such unique characteristics as wide spatial coverage, the ability to capture intricate details, and a comprehensive view of extensive geographic regions make large-scale RS imagery indispensable for tasks like scene reconnaissance and environmental monitoring.

While significant progress has been made in large-scale RS imagery, current methods for understanding such imagery can generally be divided into two main categories. The first approaches center around captioning or retrieving textual descriptions with entire images, aiming to grasp the broader scene context [6]-[8]. The second approaches focus on detecting or segmenting individual objects within the scene, such as buildings or vehicles [9]-[11]. However, when applied to large-scale RS imagery, both of these approaches encounter notable challenges:

(1) Image-level caption and retrieval fails to capture the rich semantic details embedded in large RS scenes. By matching entire images to textual descriptions, these methods overlook the detailed, region-specific semantics that are crucial in large-scale imagery. This lack of spatial precision prevents accurate localization of mid-scale semantic entities, such as clusters of buildings or infrastructural zones, which play a critical role in many practical applications.

(2) Object detection and segmentation methods primarily focus on individual objects, overlooking mid-scale semantic entities. While effective at recognizing individual targets like buildings or vehicles, these methods are insufficient for large-scale scenes, where the focus is on broader, semantic-level regions rather than isolated objects. In large-scale RS imagery, mid-scale semantic entities like building complexes or parking lots filled with vehicles offer a deeper and more contextually meaningful understanding than isolated objects, as the spatial relationships between these larger regions are crucial for comprehensive scene interpretation.

In response to these challenges, we introduce a novel task for large-scale RS imagery understanding: conte**X**tual r**e**ferring **Map** (XeMap), where specific referred regions mentioned in text must be precisely localized within an image. Notably, the query text is not limited to a straightforward description (e.g., "building complexes"), but may involve more complex referring expressions requiring contextual reasoning (e.g., "building complexes next to the playground"). This reasoning goes beyond recognizing individual entities, incorporating the spatial and environmental context to accurately localize references based on both the objects themselves and their broader surroundings. Unlike traditional tasks focused on whole-image understanding or object-level recognition, XeMap targets mid-scale referring entities within large scenes, enabling fine-grained cross-modal understanding. By focusing on these mid-scale semantic regions, XeMap naturally captures clusters of targets, providing valuable insights for large-scale scene

Yuxi Li and Lu Si contributed equally. Corresponding author: Lu Si, Jun Zhang.

Yuxi Li, Lu Si, Yujie Hou, Chengaung Liu, Bin Li, Hongjian Fang, and Jun Zhang were with Qiyuan Lab, Beijing 100095, China.

E-mail: uniluxli@qq.com (Yuxi Li), silu@qiyuanlab.com (Lu Si), 202431081015@mail.bnu.edu.cn (Yujie Hou), liuchenguang@buaa.edu.cn (Chengaung Liu), lb1037435855@163.com (Bin Li), fanghongjian@qiyuanlab.com (Hongjian Fang), zhangjun@qiyuanlab.com (Jun Zhang).

interpretation. This task bridges the gap between image-level scene interpretation, which often overlooks detailed region-specific semantics, and object-level detection, which fails to capture mid-scale semantic entities, thereby providing a new framework for comprehensively understanding the complex, information-rich content embedded in large-scale RS images. Fig. 1 illustrates the XeMap task and its relationship to captioning and retrieval, as well as detection and segmentation.

The task of XeMap is most closely related to semantic localization (SeLo) [12],[13], which refers to the task of obtaining the most relevant locations in large-scale RS images using semantic information such as text. However, SeLo focuses only on general semantic matching and cannot handle referring expressions that require specific references within the text. It identifies areas that align with the overall semantics of the query but lacks the ability to interpret more complex, context-dependent descriptions. Therefore, SeLo is limited to one-hop referring situations. In contrast, XeMap can handle multi-hop referring, matching not only semantic content but also addressing more intricate descriptions involving complex reasoning, such as "building complexes next to the playground."

In this paper, we present the first practical solution to the task of XeMap: XeMap-Network. This novel network architecture is designed to address the task of contextual localization of multi-hop referred regions in large-scale RS imagery, where a Hierarchical Multi-Scale Semantic Alignment (HMSA) module is proposed such that multi-scale visual features are aligned with the text semantic feature to facilitate precise pixel-level matching across large RS scenes. Specifically, the architecture captures detailed visual features at multiple scales, enabling accurate localization of text-referred regions (Section 3).

To overcome the lack of datasets for XeMap, we introduce XeMap-Set (Section 4), a new dataset providing pixel-level annotations for referred regions in RS imagery. XeMap-Set fills a critical gap by offering the first dataset designed specifically for training and evaluating models on this task. Given the challenges of directly annotating for the XeMap task, we propose the Multiscale Correlation Map Generation (MCMG) method, which automatically converts polygon annotations into XeMap-compatible annotations.

In summary, the main contributions of our work are as follows:

(1) A novel task: We introduce a new task, XeMap, which focuses on the pixel-level localization of multi-hop text-referred regions within large-scale RS imagery. This task goes beyond traditional image-level matching, object-level recognition, and SeLo, enabling fine-grained localization in complex and vast environments.

(2) A novel network: We introduce XeMap-Network, a network architecture specifically designed for XeMap task in large RS scenes. The fusion layer in XeMap-Network integrates multi-scale feature extraction with cross-modal attention mechanisms to ensure effective mutual information exchange. The HMSA module aligns multi-scale visual features with the text semantic vector, enabling precise multimodal matching in large-scale RS imagery.

(3) A dataset: We introduce XeMap-Set, the first dataset to provide contextual referring annotations for large remote sensing scenes, addressing the existing gap in data resources for XeMap task. MCMG method is introduced, which automatically converts polygon annotations into XeMap-compatible annotations.

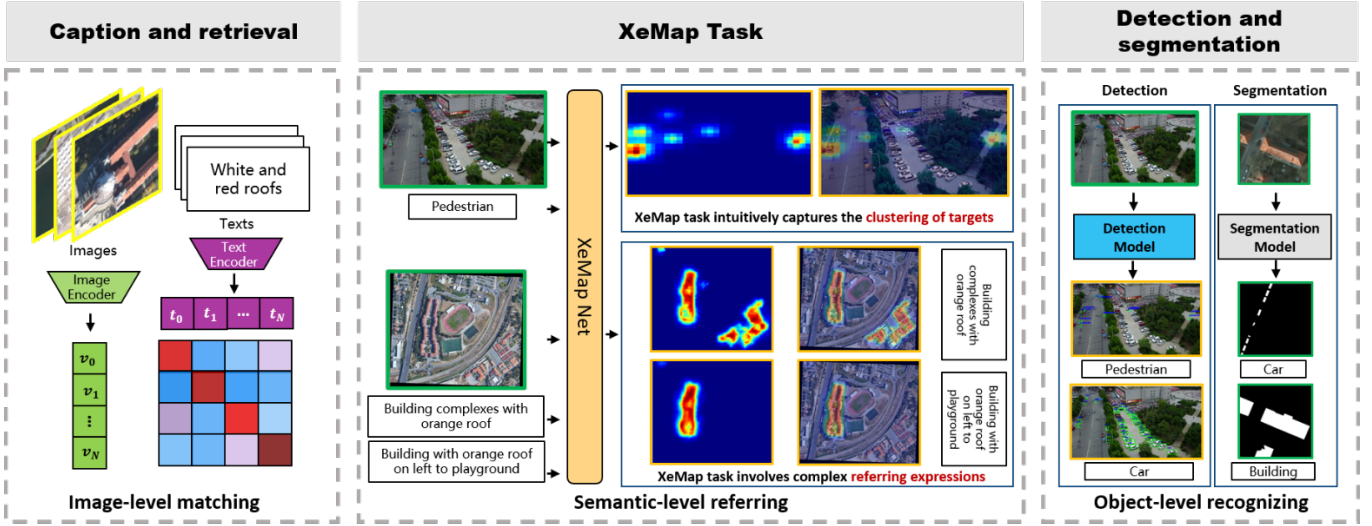

**Fig. 1.** Comparison between the XeMap task, image-level matching, and object-level recognition. Left: In captioning and retrieval, entire images are matched to textual descriptions, often overlooking the region-specific semantics critical for large-scale imagery. Right: Detection and segmentation focus on recognizing isolated objects, while semantic-level regions can sometimes provide more informative insights for large-scale scenes. Middle: The XeMap task, where specific regions referred to in the text must be precisely localized within the image. XeMap intuitively captures clusters of targets, providing valuable information for large-scale scenes. It also handles a wide range of complex referring expressions, from straightforward descriptions to more contextually rich referring expressions that require sophisticated reasoning across various spatial relationships.

## II. RELATED WORKS

*A. Cross-modal RS image retrieval*

Cross-modal RS image retrieval has recently garnered significant attention. While some research focuses on cross-modal RS image retrieval using other modalities, such as image-image retrieval [14],[15] or audio-image retrieval [16], our focus is primarily on cross-modal text-image retrieval methods. These methods can be broadly classified into two categories based on their implementation: deep neural network-based methods and those leveraging large pre-trained models.

Deep neural network-based methods often use traditional architectures like CNNs and RNNs to capture relationships between text and images, focusing on effective feature representation. Abdullah et al. [17] introduced a deep bidirectional triplet network with an LSTM as the text encoder and a pre-trained CNN as the image encoder for text-to-image matching. Rahhal et al. [18] proposed an unsupervised text-image retrieval method for RS, using a Bi-LSTM for text encoding and a pre-trained BiT model for images, optimized with an unsupervised embedding loss. Later, Rahhal et al. [19] introduced a transformer-based multilingual framework with a bidirectional contrastive loss, achieving significant retrieval improvements. Cheng et al. [20] developed a cross-modal retrieval network with a semantic alignment module incorporating attention and gate mechanisms, achieving state-of-the-art results. Yuan et al. [21] proposed LW-MCR, a lightweight retrieval model that reduces computation while maintaining performance, leveraging multi-scale information and knowledge distillation. Yuan et al. [22] further introduced GaLR, combining global and local features through dynamic fusion, with a multivariate re-rank algorithm for enhanced retrieval, also achieving state-of-the-art performance.

Methods leveraging large pre-trained models employ transformer architectures trained on vast datasets, enabling cross-modal retrieval with minimal fine-tuning. In the seminal work on CLIP [23], a two-tower model was trained to contrastively align the representations of a large number of image-text pairs sourced from the internet. Inspired by CLIP, Liu et al. proposed RemoteCLIP [24], a vision-language foundation model for RS that learns visual features and aligned text embeddings. RemoteCLIP outperforms state-of-the-art models in zero-shot image-text retrieval. Zhang et al. introduced GeoRSCLIP [25], a fine-tuned version of CLIP for RS, achieving a 3-6% improvement in cross-modal text-image retrieval using their newly created RS5M image-text paired dataset.

Despite advancements in cross-modal RS image retrieval, existing methods focus on global, image-level understanding and often overlook the region-specific semantics crucial for accurately interpreting large-scale scenes. This lack of spatial precision in localizing mid-scale entities highlights the need for methods that capture these entities and their relationships within RS imagery.

*B. Object detection and segmentation for RS imagery*

Object detection is a key task in RS, involving the identification of object instances through bounding boxes and class labels. Over the past decade, significant research has been dedicated to this area, including two-stage methods like the Fast RCNN series [26], [27], one-stage methods such as the well-known YOLO series [28]-[30], and more recent DETR variants [31]-[33] based on transformer architectures. Object detection in RS, particularly for challenges like oriented object detection, few-shot object detection, and visual grounding, remains active and rapidly evolving research directions. For oriented object detection, Xu et al. [34] proposed a framework for multi-oriented object detection by gliding the vertices of horizontal bounding boxes and introducing an obliquity factor for handling near-horizontal objects. Guo et al. [35] proposed a convex-hull feature adaptation method to enhance detection of oriented and densely packed objects by addressing spatial feature aliasing through optimized feature assignment using convex intersection over union. For few-shot object detection, Li et al. [36] proposed a meta-learning method for few-shot object detection in remote sensing images, built on the YOLOv3 architecture. Lu et al. proposed TEMO [37], a few-shot object detection method for remote sensing images that integrates text-modal knowledge to enhance classification for novel classes. For visual grounding, Sun et al. [38] introduced the task of visual grounding in remote sensing images and created the RSVG dataset. They proposed GeoVG, a method using a language encoder, image encoder, and fusion module to handle geospatial relations and large-scale scenes.

Segmentation is another rapidly evolving research direction in RS imagery processing. Two particularly emerging and pioneering directions are few-/zero-shot semantic segmentation and referring image segmentation. Jiang et al. [39] introduced the first few-shot learning method for RS segmentation, utilizing CNN-extracted features and prototype matching to label unseen object categories with minimal samples, optimized by a metric learning-based loss. Lang et al. [40] proposed R2Net, a few-shot segmentation framework that uses global rectification and decoupled registration to enhance object localization and reduce segmentation errors. Yuan et al. [41] introduced Referring Remote Sensing Image Segmentation (RRSIS) with the RefSegRS dataset and proposed a language-guided cross-scale enhancement module to improve segmentation by incorporating linguistic cues. Liu et al. [42] proposed the rotated multi-scale interaction network to address challenges in spatial scale variations and object orientations. Their model incorporates an intra-scale interaction module for fine-grained segmentation, a cross-scale interaction module for feature integration, and an adaptive rotated convolution to handle diverse orientations, significantly improving segmentation accuracy. Additionally, they developed the RRSIS-D dataset, specifically designed for RRSIS tasks.

While existing object detection and segmentation methods have advanced significantly, they often fall short in capturing the mid-scale semantic entities crucial for interpreting large-scale remote sensing imagery. Our proposed XeMap task addresses this gap by introducing pixel-level contextual reasoning, enabling a deeper understanding of complex, mid-scale relationships within large-scale scenes.

*C. Datasets for large-scale RS scenes*

Existing remote sensing (RS) datasets support tasks like image retrieval, captioning, object detection, and segmentation.

For image retrieval and captioning, the Sydney-Captions dataset [43], [44] contains 613 images, each annotated with five captions, while the RSICD dataset [45] includes 10,921 images from 30 scenes, each paired with five descriptions. The RSITMD [46] dataset provides 4,743 images and 23,715 captions for cross-modal text-image retrieval. The larger RS5M dataset [25] contains 5 million images with English descriptions for tasks like cross-modal retrieval and zero-shot classification.

For object detection, the VisDrone dataset [47] comprises 8,599 drone-captured images with over 540,000 bounding boxes, focused on pedestrian and vehicle detection. The DIOR-R dataset [48] offers 11,738 images and over 190,000 instances across 20 categories, and the DOTA v2.0 dataset [49] includes 11,268 images across 18 categories, with over 1.7 million object instances,

For segmentation, the LoveDA dataset [50] features 5,987 high-resolution images with 166,000 annotated objects, emphasizing land-cover segmentation and domain adaptation. The Vaihingen and Potsdam datasets [51] include 33 image patches with true orthophotos and digital surface models at 9 cm resolution, while the Inria Aerial Image dataset [52] contains 360 high-resolution images (3000 × 3000 pixels) over 810 km², designed for urban semantic segmentation.

While existing RS datasets support tasks such as image retrieval, captioning, object detection, and segmentation, there is currently no dataset tailored to the XeMap task. XeMap requires a dataset with annotations that specifically focus on referring expressions and the matching of mid-scale entities within large-scale RS imagery. Such a dataset would need to provide detailed labels linking textual descriptions to mid-scale objects, offering a more nuanced understanding and localization of these objects within vast RS scenes.

## III. XEMAP-NETWORK: AN END-TO-END SOLUTION FOR THE XEMAP TASK

*A. Problem Statement*

The XeMap task requires that the input query text $t_i$ be precisely localized within the given image $I_i$. The output $X_i$ is a correlation map, where higher values indicate stronger matches between the text and image pixels, and vice versa. The query text is not limited to simple descriptions but may include complex referring expressions that demand contextual reasoning. The task can be formulated as:

$$X_i = f_{\boldsymbol{\theta}}(t_i, I_i) \qquad (1)$$

where $f$ denotes the model responsible for the task, $\boldsymbol{\theta}$ represents the model parameters. Given the annotations $\Omega = \{t_i, I_i, \hat{X}_i\}_{i=1}^N$, the XeMap task corresponds to optimizing $\boldsymbol{\theta}$, such that the predicted correlation map $X_i$ is as close as possible to the annotation $\hat{X}_i$.

*B. Model Architecture*

The overall architecture of the proposed XeMap-Network is depicted in Fig. 2. It consists of three main components: an encoder layer that transforms both texts and images into tokens, a fusion layer that enhances the interaction between text and image tokens through attention mechanisms, and HMSA module. In the HMSA module, multi-scale visual features are aligned with the text semantic vector, producing the final output

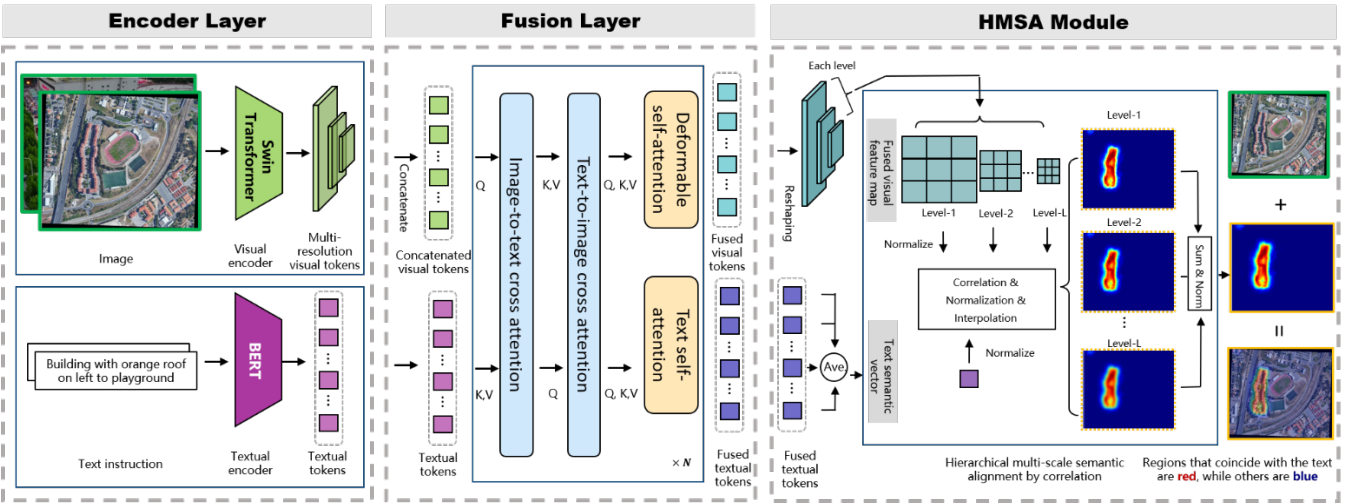

**Fig. 2.** XeMap-Network. Left: In the Encoder Layer, visual features are extracted from images using the Swin Transformer, while textual features are obtained from the BERT model. Middle: The Fusion Layer integrates cross-modal attention mechanisms, aligning text and image features through cross-attention, along with deformable self-attention and text self-attention. Right: In the Hierarchical Multi-Scale Semantic Alignment (HMSA) Module, multi-scale visual features are aligned with the text semantic vector. The resulting output highlights regions in red that correspond to the text description, while non-matching regions are shown in blue. Ave. corresponds to average operation.

by matching the textual and visual information.

**Encoder Layer.** Given an image-text pair, the text embeddings $w_i = TextEncoder(t_i)$ are extracted using a pre-trained text backbone such as BERT [53], where $w_i \in \mathbb{R}^{S \times D}$, $S$ and $D$ denotes the sequence of length and embedding dimension, respectively. For the image, a pre-trained backbone like Swin Transformer [54] is used to extract multiscale image embeddings $h_i^l = ImageEncoder(I_i)$, where $l$ refers to the multiscale feature level. The embeddings $h_i^l \in \mathbb{R}^{P_l \times D}$, with $P_l$ representing the sequence length of the level $l$ embedding features and $D$ denoting the embedding dimension. $P_l = H_l \times W_l$, where $H_l$ and $W_l$ are horizontal and vertical resolutions for embeddings at level $l$, respectively. In the implementation, the number of multiscale image levels is set to 4.

**Fusion Layer.** Since the pre-trained text and image backbones are trained independently, the fusion layer is designed to align embeddings from both modalities. To achieve this, an image-to-text and text-to-image cross-attention mechanism is applied, facilitating mutual information exchange and allowing features from both modalities to be effectively aggregated [33]. This is followed by a self-attention module applied to both image and text embeddings. Notably, a deformable self-attention module is introduced for the image embeddings, enhancing multiscale feature aggregation without relying on traditional FPN architectures [32].

We denote the fusion-enhanced image embeddings as $h_i' \in \mathbb{R}^{P \times D}$ and the fusion-enhanced text embeddings as $w_i' \in \mathbb{R}^{S \times D}$. The shape of $h_i'$ is adjusted by rearranging the feature vectors from each spatial location into a sequence. Feature maps from different scales are then concatenated to form a unified representation, resulting in a combined feature with a total sequence length of $P = \sum_l P_l$.

*C. Hierarchical Multi-Scale Semantic Alignment Module*

In the HMSA module, multi-scale visual features are aligned with the text semantic vector. The HMSA module comprises three blocks: a preprocessing block, a correlation assignment block, and a multiscale integration block.

**Preprocessing Block.** The fusion-enhanced image embeddings $w_i' \in \mathbb{R}^{S \times D}$ are averaged to generate a text semantic vector $s_i \in \mathbb{R}^D$. This vector encapsulates semantic meanings from both text and image, making it an image-aware text feature. As such, it can be treated as a query vector to examine correlations with image locations. The fusion-enhanced image embeddings $h_i'$ are reshaped back to their original multiscale feature form $g_i^l \in \mathbb{R}^{P_l \times D}$, still carrying integrated semantic meanings from the text, thus becoming a text-aware fused image feature map that acts as key vectors to be examined by the text semantic vector.

**Correlation Assignment Block.** Firstly, both the text semantic vector $s_i$ and the text-aware fused image feature map $g_i^l$ are normalized so that the norm of each vector equals one. Then, the correlation is computed between $s_i$ and $g_i^l$ from each level:

$$c_i^l(p,q) = g_i^l(p,q) \cdot s_i \qquad (2)$$

where $(p,q)$ corresponds to a 2D position on the feature map.

Finally, we normalize the correlation output into range of [0-1] using $\tilde{c}_i^l = (c_i^l + 1)/2$, and apply bilinear interpolation to up-sample the resolution to match the original image size.

**Multiscale Integration Block.** The final output correlation map is generated by summing the sub-correlation maps from each level and averaging them across the total number of levels, as shown below

$$x_i(p,q) = \frac{1}{L}\sum_{l=1}^{L} \tilde{c}_i^l(p,q) \qquad (3)$$

where $L$ is the total number of multiscale image levels, and $(p,q)$ corresponds to a 2D position on the original image.

*D. Loss Function*

We utilize the L1 loss to measure the absolute differences between the predicted correlation map and the ground truth:

$$Loss = \sum_p \sum_q \left| \hat{X}_i(p,q) - X_i(p,q) \right| \qquad (4)$$

where $\hat{X}_i(p,q)$ corresponds to the annotated correlation map, and $(p,q)$ represents a 2D position on the correlation map. This loss encourages precise pixel-level predictions, which is crucial for accurately localizing the referred regions in the XeMap task.

## IV. XeMap-Set: A Dataset for the XeMap Task on Large-Scale RS Images

This section provides a comprehensive overview of the XeMap-Set, which consists of a training-validation portion and a testing portion. We begin with an overview of the dataset, followed by an introduction to the MCMG method. Next, we outline the annotation protocol for the test partition. Finally, we present an analysis of the XeMap-Set.

*A. Overview of XeMap-Set*

XeMap-Set is built from publicly available datasets designed for various tasks: RefCOCO (for referring task), VisDrone [47], DIOR-R [48], DOTA v2.0. [49] (for objection), LoveDA [50], Vaihingen and Potsdam[51], Inria Aerial Image [52] (for segmentation), and AIR-SLT (for SeLo) [12]. However, these datasets lack XeMap-specific annotations. To bridge this gap, we propose the MCMG method, which automatically converts referring, detection, and segmentation annotations into XeMap-compatible annotations.

*B. Multiscale Correlation Map Generation Method*

The training-validation portion of XeMap-Set is generated using the MCMG method, which involves three key processes: targeted region mask generation, multiscale grid overlap analysis, and correlation map enhancement. This portion includes images from the aforementioned datasets, excluding AIR-SLT. Refer to Fig. 3 for an overview of the MCMG method.

**Targeted Region Mask Generation.** The process starts by converting annotations from referring, detection, and segmentation tasks into a targeted region mask, where pixels within the designated regions are marked as one. For detection and segmentation tasks, labels are translated into textual descriptions, and a corresponding mask is generated for each category.

**Multiscale Grid Overlap Analysis.** To handle the diverse object sizes within RS scenes, the image is divided into grids of multiple scales. The targeted region mask is then projected onto these grids, allowing for the calculation of the overlap ratio between the targeted region mask and each grid cell. The greyscale of each grid is calculated by:

$$E_l(p,q) = \frac{A_l}{\text{Intersect}[G_l(p,q), M]} \quad (5)$$

where $(p,q)$ corresponds to a 2D position on the grid, $A_l$ denotes the area of each grid with level $l$, $G_l(p,q)$ represents the generated grid with position $(p,q)$, $\text{Intersect}[G_l(p,q), M]$ denotes calculating the intersection area of grid $G_l(p,q)$ and the targeted region mask $M$.

This step quantifies the extent to which objects occupy the grid cells, capturing object density across different spatial scales and ensuring accurate representation of both small and large objects. The results are aggregated to generate a comprehensive multiscale representation.

**Correlation Map Enhancement.** This process involves two key sub-processes: Gaussian smoothing and normalization. First, the aggregated overlap ratios undergo Gaussian smoothing to create a soft probability map, enhancing the spatial continuity of the correlation map for a refined representation. Next, the correlation map is normalized to a [0-255] range, ensuring the output is suitable for visualization and further processing.

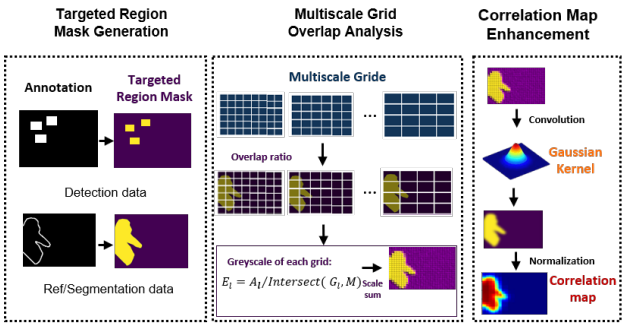

**Fig. 3.** Multiscale correlation map generation method.

*C. Annotation Protocol for Test Partition*

We provide a test set specifically for the XeMap task, comprising images sourced from the test partitions of VisDrone [47], DIOR-R [48], DOTA v2.0. [49], LoveDA [50], Inria Aerial Image [52], and AIR-SLT [12]. Each sample includes an RS image with a resolution ranging from 1k × 1k to 10k × 10k, a textual query, and one or more corresponding polygon regions. Several examples of the annotated data can be seen from the first column of Fig. 5.

**Annotation Protocol.** For labeling, we adopted a protocol similar to that used in ReferItGame [55] and GigaGrounding [56], involving two groups of annotators. The first group generated the textual query and its corresponding polygon annotation, while the second group was tasked with drawing bounding boxes based solely on the provided query. A sample was considered valid if the Intersection over Union (IoU) between the two groups' annotations was greater than 0.5.

In the annotation process, annotators were provided with reference categories for suggested target regions, including clusters of people, vehicles, residential buildings, as well as aggregated vegetation, parking lots, aircraft parking areas, roads, and various others not exhaustively listed here. These target regions needed to be clearly visible and occupy between 0.2% and 75% of the total image area.

Two types of attributes were recorded during annotation: multi-hop expressions and multi-ref expressions. A multi-hop expression requires the model to identify additional reference objects as part of the localization process. For instance, "building next to the playground" is a multi-hop expression, as the model must first locate the playground before finding the building next to it. A multi-ref expression, on the other hand, involves multiple regions that match the query. For example, "parking lots filled with cars" may be a multi-ref expression, as there may be several parking lots that satisfy the description.

**Test Metric.** Following SeLo [12], we utilize the following metrics to evaluate the correlation map against the annotated polygon regions: significant area proportion ($R_{su}$), attention shift distance ($R_{as}$), discrete attention distance ($R_{da}$), and the unified metric ($R_{mi}$). $R_{su}$ measures the proportion of attention focused on the ground-truth area compared to non-ground-truth regions. $R_{as}$ quantifies the distance by which attention shifts from the ground-truth center. $R_{da}$ evaluates the distribution of generated attention based on probability divergence and the number of candidate attention points. $R_{mi}$ provides a comprehensive metric encompassing all these evaluations.

*D. Analysis of XeMap-Set*

This section provides an analysis of the key characteristics of XeMap-Set from following perspectives.

**Data Sources.** XeMap-Set comprises images from diverse sources. Table 1 summarizes the data origins, categorizing them into Refer data, RS data, and SeLo data. It details the number of training, validation, and test samples for each source, illustrating the dataset's diversity and comprehensiveness for training and evaluation.

**Statistics for Test Partition.** We annotated 300 images for the test partition, resulting in 1,129 queries, averaging 3.76 queries per image. The number of images sourced from each dataset is listed in Table 1. The distribution of the XeMap-Set test partition is presented in Fig. 4, which illustrates various aspects of the data, including query counts, multi-hop counts, multi-ref counts relative to image counts, and expression length relative to query counts. The analysis shows that images with 3 to 4 queries are the most frequent, and similarly, those containing 2 multi-hop expressions and 1 multi-ref expression are the most common. The majority of expressions are between 5 to 10 words in length, with a notable number ranging from 10 to 15 words.

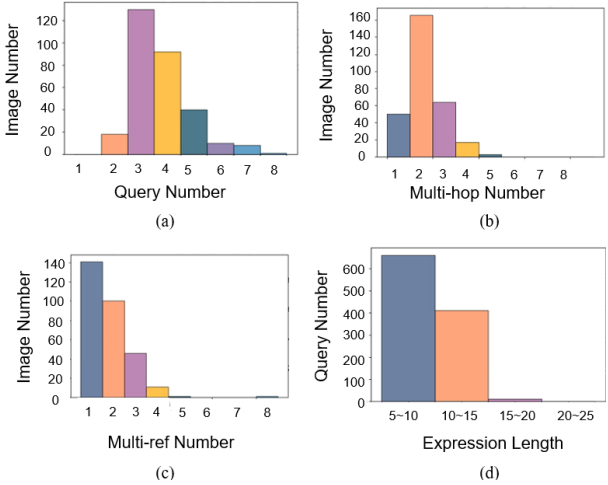

**Fig. 4.** Histograms representing the distribution of the XeMap-Set test partition. (a) Number of images categorized by varying query counts. (b) Number of images categorized by the count of multi-hop expressions. (c) Number of images categorized by the count of multi-ref expressions. (d) Number of queries categorized by different expression lengths.

TABLE 1
SUMMARY OF XEMAP-SET DATA SOURCES

|  | Source | Train/Val | Test |
|---|---|---|---|
| Refer Data | RefCOCO | 49676/13205 | 0 |
| RS Data | VisDrone | 6471/548 | 42 |
|  | DIOR | 8000/3725 | 51 |
|  | DOTAv2 | 1830/593 | 49 |
|  | LoveDa | 3577/614 | 100 |
|  | Potsdam-Vaihingen | 2738/606 | 0 |
|  | AerialImage | 14418/162 | 36 |
| SeLo Data | AIR-SLT | 0/0 | 22 |
| Summary | Images | 86710/19453 | 300 |
|  | Queries | 791971/84576 | 1129 |
|  | Queries per Image | 9.13/4.35 | 3.76 |

## V. EXPERIMENTS AND EVALUATIONS

In this section, we showcase the effectiveness of the proposed XeMap-Network by pretraining it on the training partition of XeMap-Set and evaluating its zero-shot performance on the annotated test partition of XeMap-Set.

### A. Implementation Details

The XeMap-Network is built upon the MMDetection toolbox, utilizing open-source BERT [53] and Swin Transformer (Swin-T) [54] as the text and image encoders. During training, only the parameters in the fusion layers are tuned, using the training partition of XeMap-Set for 25 epochs. The AdamW optimizer is employed with a learning rate of $1e^{-3}$ and a weight decay of $1e^{-4}$. Images and annotations are resized to 512×512, and the batch size is set to 400. Training takes approximately 12 hours on 10×RTX 3090 GPUs.

### B. Zero-Shot Performance of XeMap-Network

We evaluate the performance of the XeMap-Network on the XeMap task using the test partition of XeMap-Set. Since the test partition is newly annotated and was not used during training, this evaluation can be regarded as zero-shot performance. The results are presented in Table 2, comparing XeMap-Network against SeLo [12], SeLov2 [13], CLIP-SeLo, and CLIP-SeLov2. In both CLIP-SeLo and CLIP-SeLov2, the CLIP model is fine-tuned on the RSITMD [46] dataset to enhance performance, with the final layers of the text and image encoders fine-tuned for 10 epochs. The SGD optimizer was used, with a learning rate of $1e^{-4}$ and a batch size of 500.

As shown in Table 2, XeMap-Network achieves the highest performance in the averaged unified metric ($R_{mi}$), averaged significant area proportion ($R_{su}$), averaged attention shift distance ($R_{as}$), while securing the second-best result for averaged discrete attention distance ($R_{da}$). Given that Rmi serves as a comprehensive metric encompassing all these evaluations, XeMap-Network attains the highest score of 0.5789, significantly outperforming the second-best score of 0.4903 from state-of-the-art methods. We also recorded the time consumption for each method. XeMap-Network processed 1,129 queries in 201 seconds, significantly faster than the second-best approach. The evaluation was conducted on a RTX 3090 GPU.

### C. Zero-Shot Performance of XeMap-Network

Figure 5 presents qualitative results. For straightforward queries, such as "the road running through the image" (Fig. 5, row (a)), XeMap-Network delivers the most accurate visualization. It also excels in handling more complex descriptions, such as "the T-shaped white road" (Fig. 5, row (b)), and generates a precise correlation map. Additionally, XeMap-Network demonstrates a superior grasp of spatial orientation, accurately identifying "the river in the bottom right corner" and "the factory above the image" (Fig. 5, rows (c) and (d)), outperforming other methods.

TABLE 2
ZERO-SHOT EVALUATION ON XEMAP-SET TEST PARTITION

| Model | Text Encoder | Image Encoder | Fusion Method | Pre-training Data | ↑ $R_{mi}$ | ↑ $R_{su}$ | ↓ $R_{as}$ | ↑ $R_{da}$ | ↓ Total Time (s) |
|---|---|---|---|---|---|---|---|---|---|
| | | | | | (Zero-Shot on XeMap-Set Test Partition) | | | | (1129 Queries) |
| SeLo | Gated Recurrent Unit | CNN with Multi-level Visual Self-Attention | RS Image-Text Similarity | RSITMD | 0.4275 | 0.5969 | 0.7343 | 0.3829 | 3,536 |
| SeLov2 | Gated Recurrent Unit | CNN with Multi-level Visual Self-Attention | Multilevel Likelihood | RSITMD | 0.4202 | 0.5743 | 0.7268 | 0.3796 | 2,439 |
| CLIP-SeLo | Transformer based CLIP-Text Encoder | ViT-L/14 | RS Image-Text Similarity | WebImageText + RSITMD | 0.4711 | 0.6449 | 0.6963 | 0.4275 | 10,175 |
| CLIP-SeLov2 | Transformer based CLIP-Text Encoder | ViT-L/14 | Multilevel Likelihood | WebImageText + RSITMD | 0.4903 | 0.6358 | 0.6617 | **0.4701** | 8,471 |
| XeMap-Network (Ours) | BERT | Swin-T | HMSA module | XeMap-Set Training Partition | **0.5789** | **0.8651** | 0.6606 | 0.4565 | **201** |

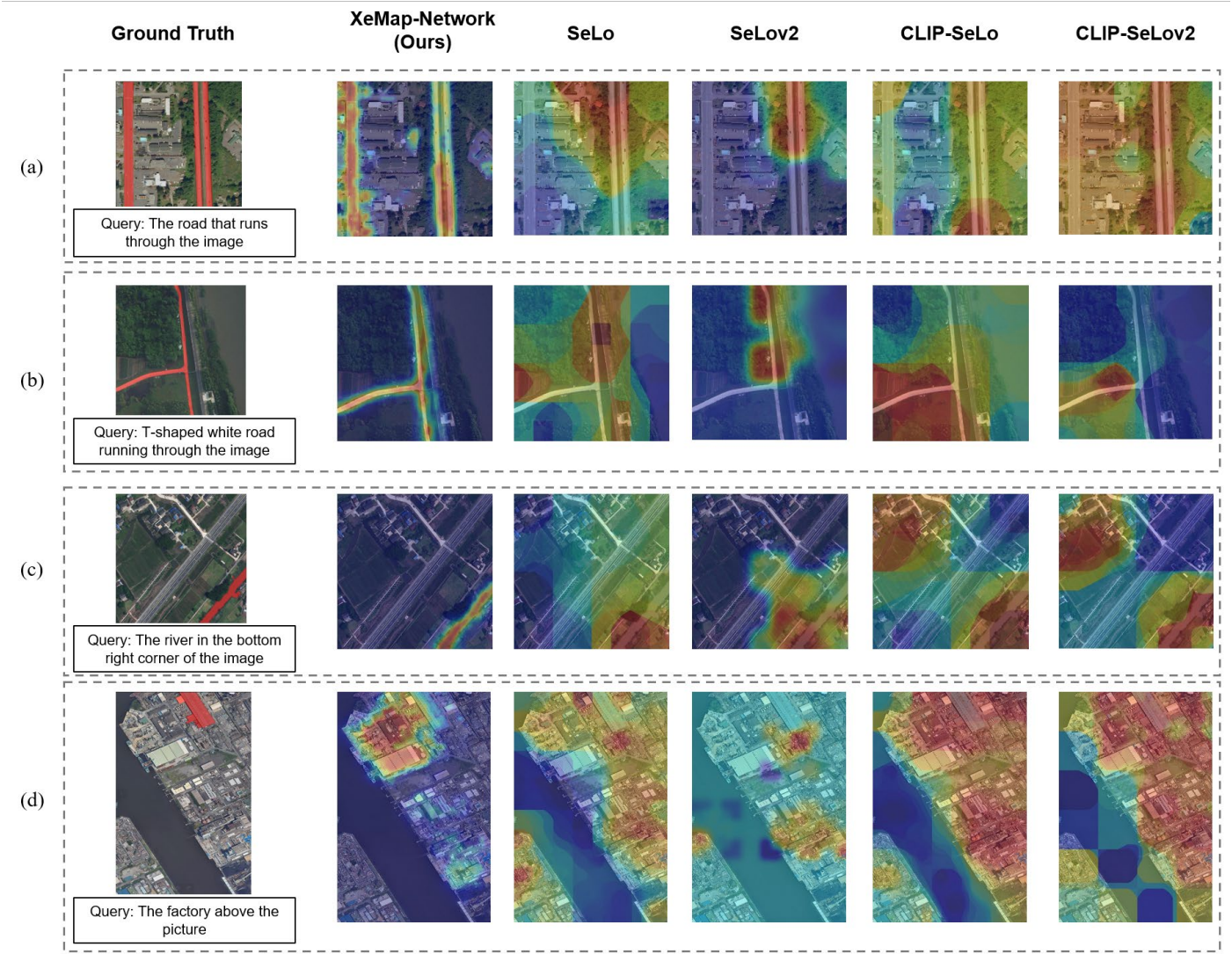

**Fig. 5.** Qualitative results comparing XeMap-Network against state-of-the-art methods. (a) Straightforward query: XeMap-Network provides the most accurate visualization. (b) Query with more complex description: XeMap-Network accurately maps the T-shaped white road, which was not encountered during training. (c) and (d) Spatial orientation queries: XeMap-Network demonstrates a superior understanding of spatial orientation, accurately identifying "the river in the bottom right corner" and "the factory above the image".

## D. Ablation study

In this section, we conduct an ablation study to evaluate different methods for generating the text semantic vector within XeMap-Network. We trained three different versions of XeMap-Network, using max pooling, selecting the first vector representing the overall meaning of the text, and average pooling (used in main text). The evaluation metrics used are the averaged unified metric ($R_{mi}$), averaged significant area proportion ($R_{su}$), averaged attention shift distance ($R_{as}$), and averaged discrete attention distance ($R_{da}$), conducted on the test partition of XeMap-Set.

Based on the ablation results presented in Table 3, we can draw the following conclusions regarding the selection of the most suitable method for generating the text semantic vector:

**Max Pooling.** This method achieved relatively good results, especially in $R_{mi}$ (0.5758), which indicates a strong overall feature extraction capability. However, it performed slightly worse in $R_{as}$ (0.6615) and $R_{da}$ (0.4454), suggesting that it might have limitations in maintaining spatial attention consistency.

**First Vector.** Selecting the first vector representing the overall meaning of the text showed the highest performance in $R_{su}$ (0.8764), suggesting that it captured significant areas effectively. However, it had the lowest $R_{mi}$ (0.5429) and higher $R_{as}$ (0.7282), indicating that while it could highlight key regions, it may lack robustness in overall performance and result in larger attention shifts.

**Average Pooling.** The average pooling method provided the best balance across all metrics. It had the highest $R_{mi}$ (0.5789), which indicates the best overall performance in feature extraction. It also maintained competitive scores in $R_{su}$ (0.8651), $R_{as}$ (0.6606), and $R_{da}$ (0.4565), demonstrating its ability to generalize well across different aspects of the test data.

The average pooling method was selected as the optimal approach for generating the text semantic vector in XeMap-Network. This choice was made because average pooling offered the highest $R_{mi}$ value, indicating the best overall feature representation, while also maintaining competitive performance in other metrics such as $R_{su}$, $R_{as}$, and $R_{da}$.

TABLE 3
ABLATION STUDY

| Text Sematic Vector Generation Method | ↑ $R_{mi}$ | ↑ $R_{su}$ | ↓ $R_{as}$ | ↑ $R_{da}$ |
|---|---|---|---|---|
| | (Zero-Shot on XeMap-Set Test Partition) | | | |
| Max Pooling | 0.5758 | 0.8649 | 0.6615 | 0.4454 |
| First Vector | 0.5429 | **0.8764** | 0.7282 | 0.3886 |
| Average Pooling | **0.5789** | 0.8651 | **0.6606** | **0.4565** |

## E. Zero-Shot Performance of XeMap-Network on AIR-SLT Dataset

In this section, we evaluate the performance of XeMap-Network on the AIR-SLT dataset against state-of-the-art methods, demonstrating the distinct objectives of the XeMap and SeLo tasks. While the SeLo task aims to locate all semantic elements related to the query within the image, XeMap focuses solely on the specific referred entity. As shown in Fig. 6:

In Fig. 6(a), with the query "There is a gray road between the green football field and the green lake", AIR-SLT annotates all related objects, including the road, green lake, and football field, whereas XeMap locates only the road between the lake and the football field.

In Fig. 6(b), with the query "A gray road is built beside a dark green lake surrounded by greenery", AIR-SLT includes both the lake and surrounding road in the annotation, while XeMap specifically targets the road surrounding the lake.

In Fig. 6(c), for the query "Two yellow cranes are unloading from a ship full of cargo", AIR-SLT annotates both the cranes and the ship, whereas XeMap only highlights the yellow cranes.

In Fig. 6(d), with the query "A white plane parked in a tawny clearing inside the airport", AIR-SLT marks both the airplane and airport, while XeMap focuses only on the airplane.

## F. Qualitative Results for Multi-hop Expressions

In this section, we qualitatively demonstrate multi-hop expressions using XeMap-Network. One image from each dataset—AIR-SLT, VisDrone, DIOR, DOTA, LoveDA, and AerialImage—is selected, and queries containing multi-hop expressions are used to require the model to generate the corresponding correlation maps. To further illustrate the model's reasoning process, we break down the multi-hop expressions into one-hop components, observing how the model generates the final response through these incremental steps. The results are presented in Fig. 7. For comparison, we also present multi-hop expressions for the CLIP-SeLov2 model in Fig. 8, which demonstrates the best performance on the XeMap-Set test partition, shown in main text. While CLIP-SeLov2 is able to partially locate one-hop expressions, it lacks sufficient precision and granularity. For multi-hop expressions, the model fails completely, indicating that CLIP-SeLov2 lacks the capability to effectively understand and process multi-hop reasoning.

## V. CONCLUSION

We introduced the XeMap task, a problem that aims at contextual localization of referred regions in large-scale RS imagery. We developed XeMap-Network, a first practical solution that effectively captures mid-scale semantic entities often overlooked by existing methods. Experimental results demonstrate the effectiveness of our approach. We hope XeMap could pave the way for further research in multimodal analysis, impacting fields such as infrared-visible RS, and enabling comprehensive insights into complex geographic and environmental data.

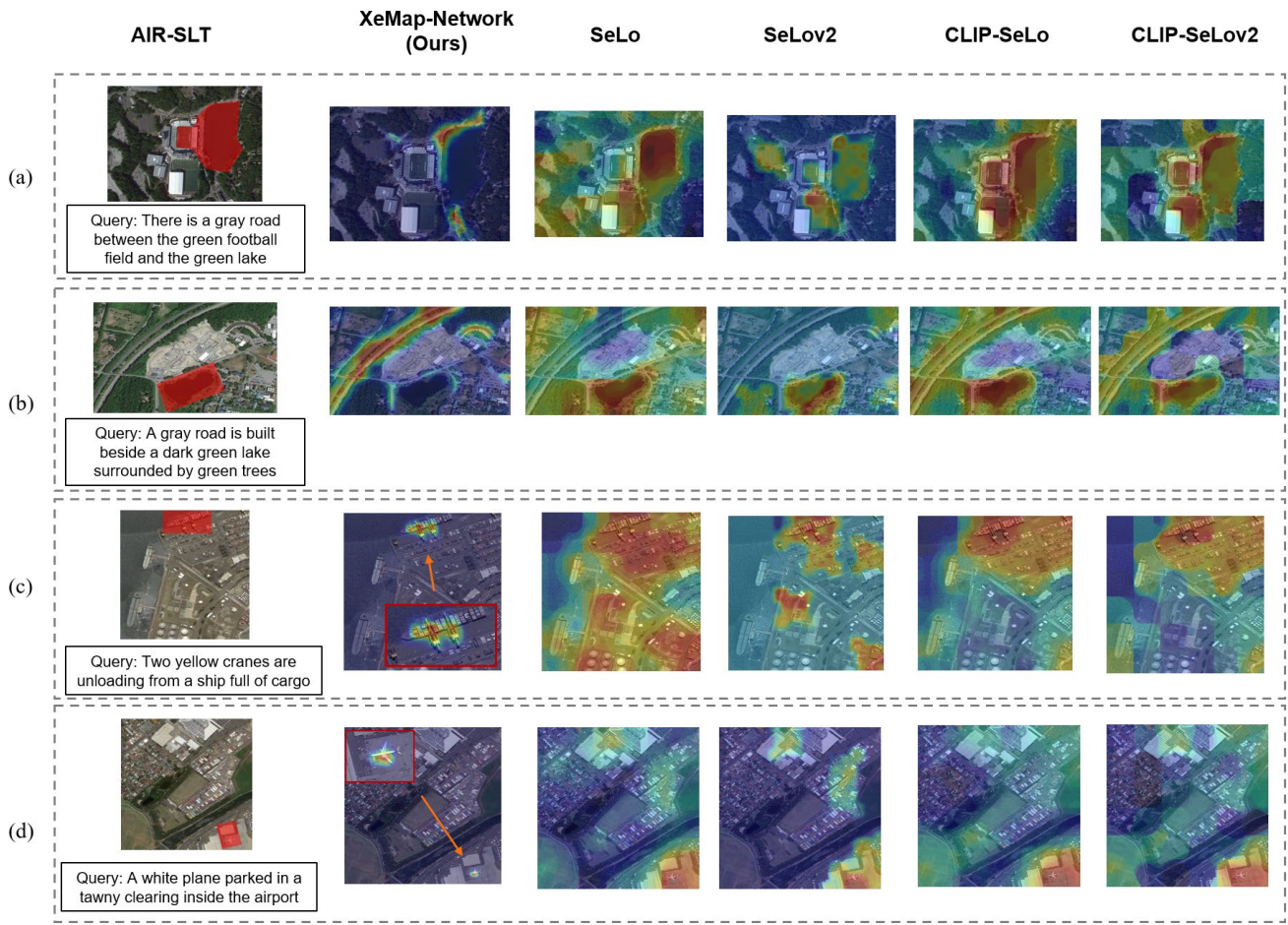

**Fig. 6.** Comparison on the AIR-SLT Dataset. This figure qualitatively demonstrates the difference between the SeLo task and the XeMap task. The SeLo task aims to locate all related semantic information within the image, while the XeMap task focuses solely on locating the specifically referred entity. (a) There is a gray road between the green football field and the green lake. (b) A grey road is built beside a dark green lake surrounded by green lake. (c) Two yellow cranes are unloading from a ship full of cargo. (d) A white plane parked in a tawny clearing inside the airport.

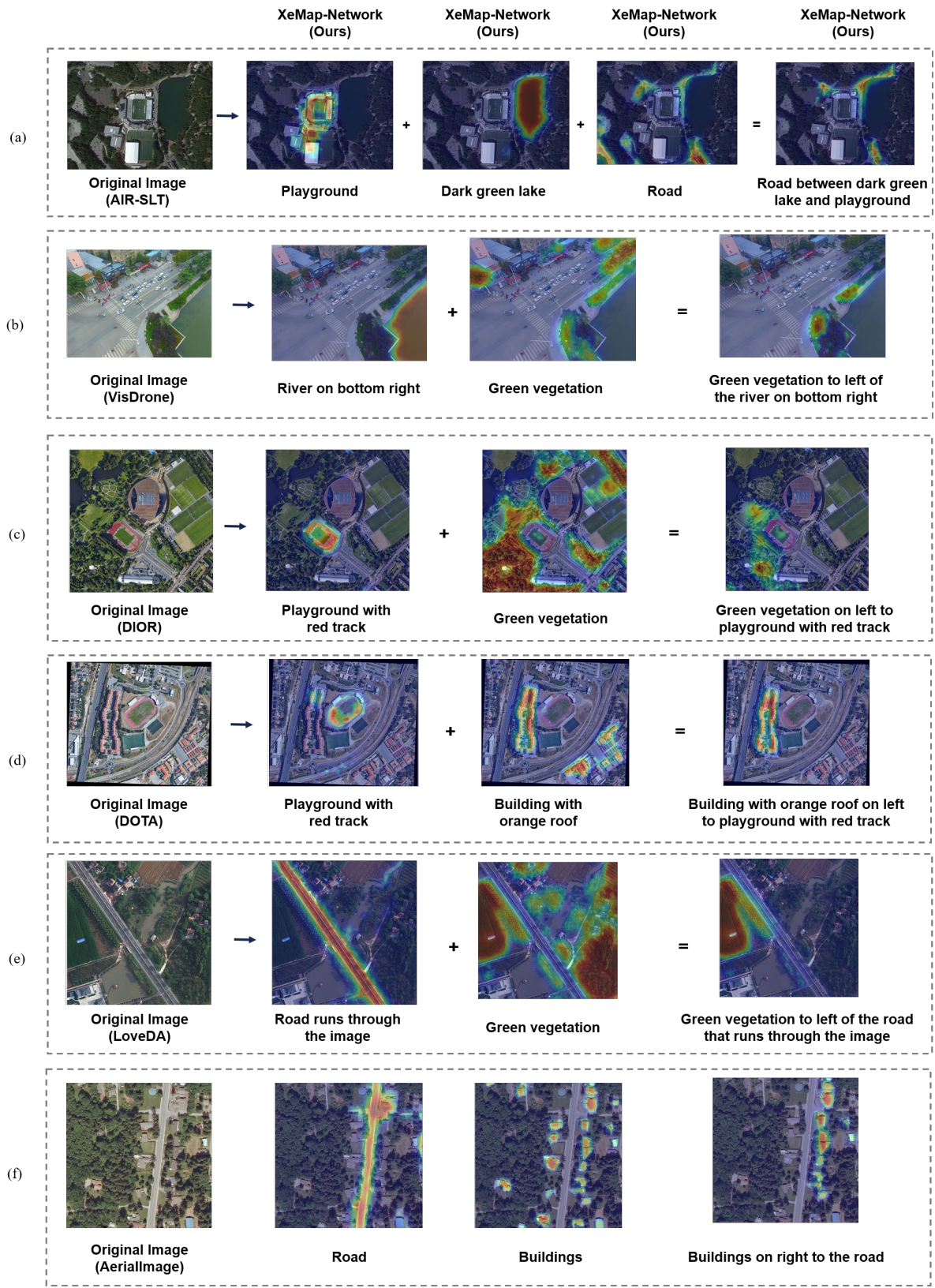

**Fig. 7.** Multi-hop expressions for XeMap-Network. This figure qualitatively demonstrates multi-hop expressions using our XeMap-Network. The first column displays the original images, with one image selected from each dataset: AIR-SLT, VisDrone, DIOR, DOTA, LoveDA, and AerialImage. The center columns represent one-hop expressions that the model initially locates, while the last column shows the results for the multi-hop expressions generated by the model.

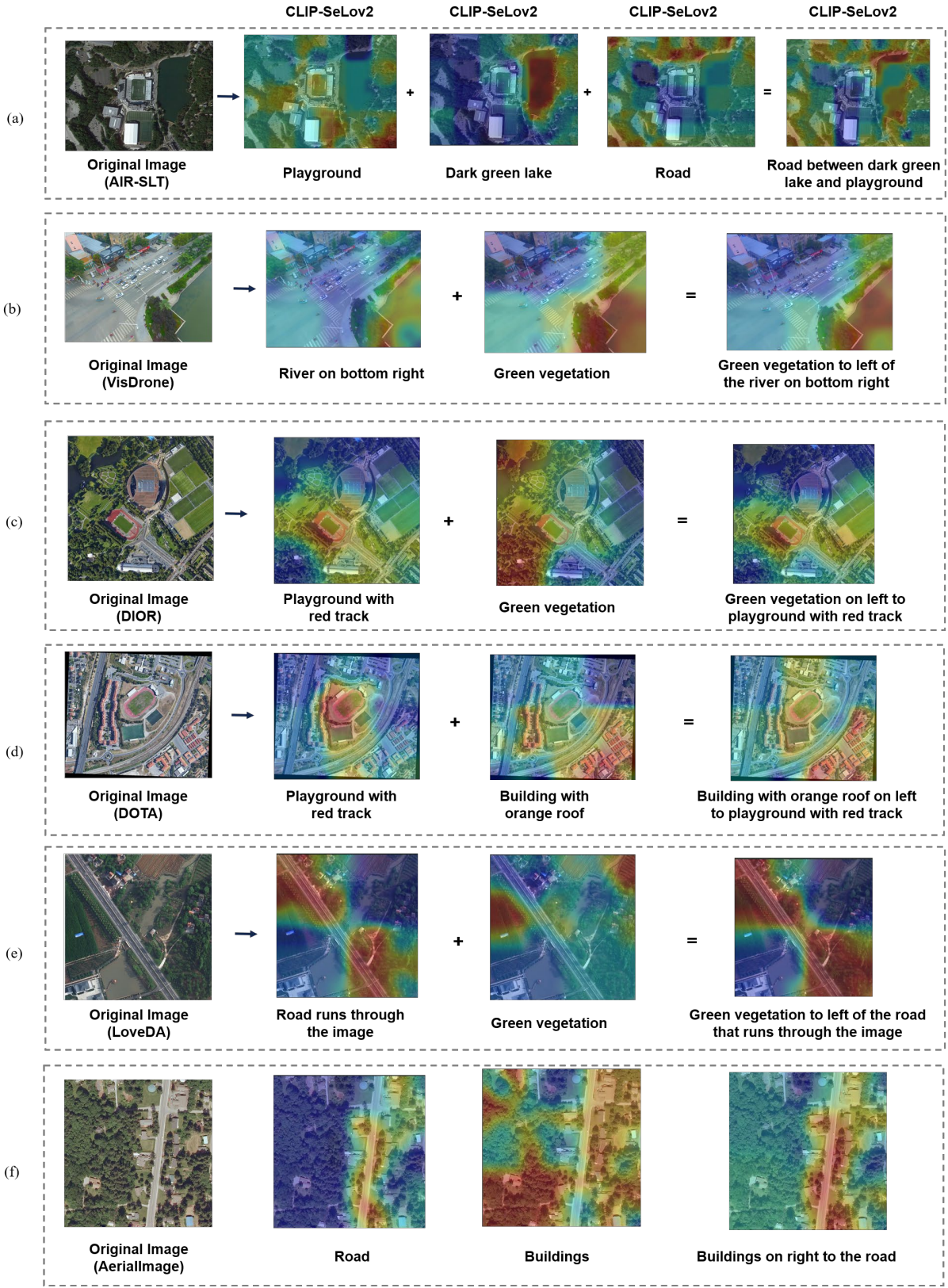

**Fig. 8.** Multi-hop expressions for CLIP-SeLov2. This figure qualitatively demonstrates that SeLo does not have ability for multi-hop expressions. The first column displays the original images, with one image selected from each dataset: AIR-SLT, VisDrone, DIOR, DOTA, LoveDA, and AerialImage. The center columns represent one-hop expressions that the model initially locates, while the last column shows the results for the multi-hop expressions generated by the model.